\documentclass[pdflatex,sn-mathphys-num]{sn-jnl}

\usepackage{graphicx}%
\usepackage{multirow}%
\usepackage{natbib}
\usepackage{amsmath,amssymb,amsfonts}%
\usepackage{amsthm}%
\usepackage{mathrsfs}%
\usepackage[title]{appendix}%
\usepackage{xcolor}%
\usepackage{textcomp}%
\usepackage{bm}
\usepackage{manyfoot}%
\usepackage{booktabs}%
\usepackage{algorithm}%
\usepackage{algorithmicx}%
\usepackage{algpseudocode}%
\usepackage{listings}%
\usepackage{siunitx}
\usepackage{setspace}
\unnumbered

\begin{document}

\title[Continuous drive heterodyne microwave sensing with spin qubits in hexagonal boron nitride]{Continuous drive heterodyne microwave sensing with spin qubits in hexagonal boron nitride}


\author[1, *]{Charlie J. Patrickson}

\author[1]{Valentin Haemmerli}

\author[1]{Shi Guo}

\author[2]{Andrew J. Ramsay}

\author[1]{Isaac J. Luxmoore}

\affil[1]{Department of Engineering, University of Exeter, EX4 4QF, UK} 

\affil[2]{Hitachi Cambridge Laboratory, Hitachi Europe Ltd.,  CB3 0HE,UK} 

\affil[*]{corresponding author: Charlie J. Patrickson (cp728@exeter.ac.uk)}


\abstract{Quantum sensors that use solid state spin defects have emerged as effective probes of weak alternating magnetic signals. By recording the phase of a signal relative to an external clock, these devices can resolve signal frequencies to a precision orders of magnitude longer than the spin state lifetime. However, these quantum heterodyne protocols suffer from sub-optimal sensitivity, as they are currently limited to pulsed spin control techniques, which are susceptible to cumulative pulse-area errors, or single continuous drives which offer no protection of the spin coherence. Here, we present a control scheme based on a continuous microwave drive that extends spin coherence towards the effective $T_2 \approx \frac{1}{2}T_1$ limit and can resolve the frequency, amplitude and phase of GHz magnetic fields. The scheme is demonstrated using an ensemble of boron vacancies in hexagonal boron nitride, and achieves an amplitude sensitivity of $\eta \approx 3-5 \:\mathrm{\mu T \sqrt{Hz}}$ and phase sensitivity of $\eta_{\phi} \approx 0.076 \:\mathrm{rads \sqrt{Hz}}$. By repeatedly referencing the phase of a resonant signal against the coherent continuous microwave drive in a quantum heterodyne demonstration, we measure a GHz signal with a resolution $<$1 Hz over a 10 s measurement. Achieving this level of performance in a two-dimensional material platform could have broad applications, from probing nanoscale condensed matter systems to integration into heterostructures for quantum networking.}

\maketitle

                              

\section*{Introduction}\label{sec:level1}

Electron spins confined to crystalline point defects have attracted significant interest for magnetic field sensing in ambient conditions. To date,  progress in the field has been dominated by the nitrogen vacancy (NV) center in diamond, with sustained efforts in material engineering and measurement protocols producing notable applications in materials science\cite{Bertelli2020, Zhou2020}, biosensing\cite{Arai2022, PhysRevApplied2021_Cao} and nanoscale NMR \cite{Staudacher2013}. For the detection of AC fields, outstanding amplitude sensitivities in the $\mathrm{pT/\sqrt{Hz}}$ range\cite{Wang2022, alsid_2023} have been demonstrated by using large ensembles of NV-centers embedded in bulk material. However, achieving comparable performance in small nanodiamonds, where the sensor can be placed in close proximity to a signal source, has remained a challenge \cite{Doherty_PhysRep2013, Alkahtani2019, Cao2020}. This has motivated growing interest in alternative materials, with two-dimensional foils of hexagonal boron nitride showing particular promise\cite{Gottscholl2021, Vaidya2023}. The system hosts multiple optically active defects capable of coherent spin control, most notably the boron vacancy\cite{Gottscholl2020}, and recently discovered single carbon-related defects of ambiguous structure\cite{Stern2022, Stern2024}.

Two important metrics in the characterisation of spin-based sensors are the frequency resolution and the sensitivity to low amplitude signals. Effective sensing protocols enhance both properties, which is often achieved by extending the spin state lifetime using dynamical decoupling. For NV centres, this has lead to coherence times approaching an effective limit of $T_2 \approx \frac{1}{2}T_1$ \cite{BarGill2013}, with the frequency resolution and sensitivity similarly bounded. However, by exploiting the spin's sensitivity to signal phase, pulsed quantum heterodyne measurements have achieved frequency resolutions far beyond the limit imposed by $T_1$ for signal frequencies $<$ 50 MHz\cite{Boss_Science_2017, Schmitt_Science_2017, Glenn2018, Jiang2023}. The technique builds up a stroboscopic image of a signal waveform by recording the instantaneous signal phase against an external clock, which can have a coherence many orders of magnitude better than $T_1$. To access GHz frequencies, other heterodyne approaches have used a pulsed microwave (MW) drive that dresses the electron spin transition with additional energy levels, which form a Mollow triplet in the absorption spectrum\cite{Joas2017, Meinel2021}. However pulsed methods are sub-optimal, as the sensitivity suffers from pulse-area errors\cite{Lang2019, Ishikawa2018} and requires high microwave powers\cite{Aharon2019}. Alternatively, single continuous drives have been used to replicate the phase response, \cite{Staudenmaier2021, Meinel2021}, but the sensitivity of these schemes is limited by coherence times $<<T_2$.

In this work, we present a continuous drive sensing protocol that can detect the phase, frequency and amplitude of an AC magnetic field, whilst also achieving coherence times approaching $T_2 \approx \frac{1}{2}T_1$. The scheme uses a single, continuous phase modulated microwave field, often referred to as continuous concatenated dynamical decoupling (CCDD)\cite{Cai2012, Farfurnik2017}, to drive the spin along two different axes, at two different frequencies. A signal that is resonant with these rotations will cause the spin to deviate from the CCDD driven trajectory, with a path that depends on the signal phase. Information about the new trajectory, and therefore signal phase, is revealed by projecting the spin onto the z-axis through optical readout. We demonstrate the scheme using a $V_B^-$ ensemble in hexagonal boron nitride, where it successfully suppresses the effects of magnetic noise from the host III-V nuclear spin bath. In comparison to other CCDD sensing schemes\cite{Stark2017}, we show that the microwave drive parameters can be used to switch the spin response between phase and amplitude detection. The device's amplitude and phase sensitivity are quantified as, $\eta \approx 3-5 ~\mathrm{\mu T \sqrt{Hz}}$ and $\eta_{\phi} \approx 0.076 ~\mathrm{rads \sqrt{Hz}}$, respectively. Finally, we use the scheme to repeatedly record signal phase in a quantum heterodyne protocol, achieving a frequency resolution of 0.118 Hz at $\sim$2.31 GHz, over a total measurement time of 10 seconds. The scheme provides comprehensive characterisation of AC signals, without the accumulation of errors found in pulsed sequences, and is equally applicable to trapped atom and ion systems, and other solid state defects.

\begin{figure*}
\includegraphics[width=1\columnwidth]{{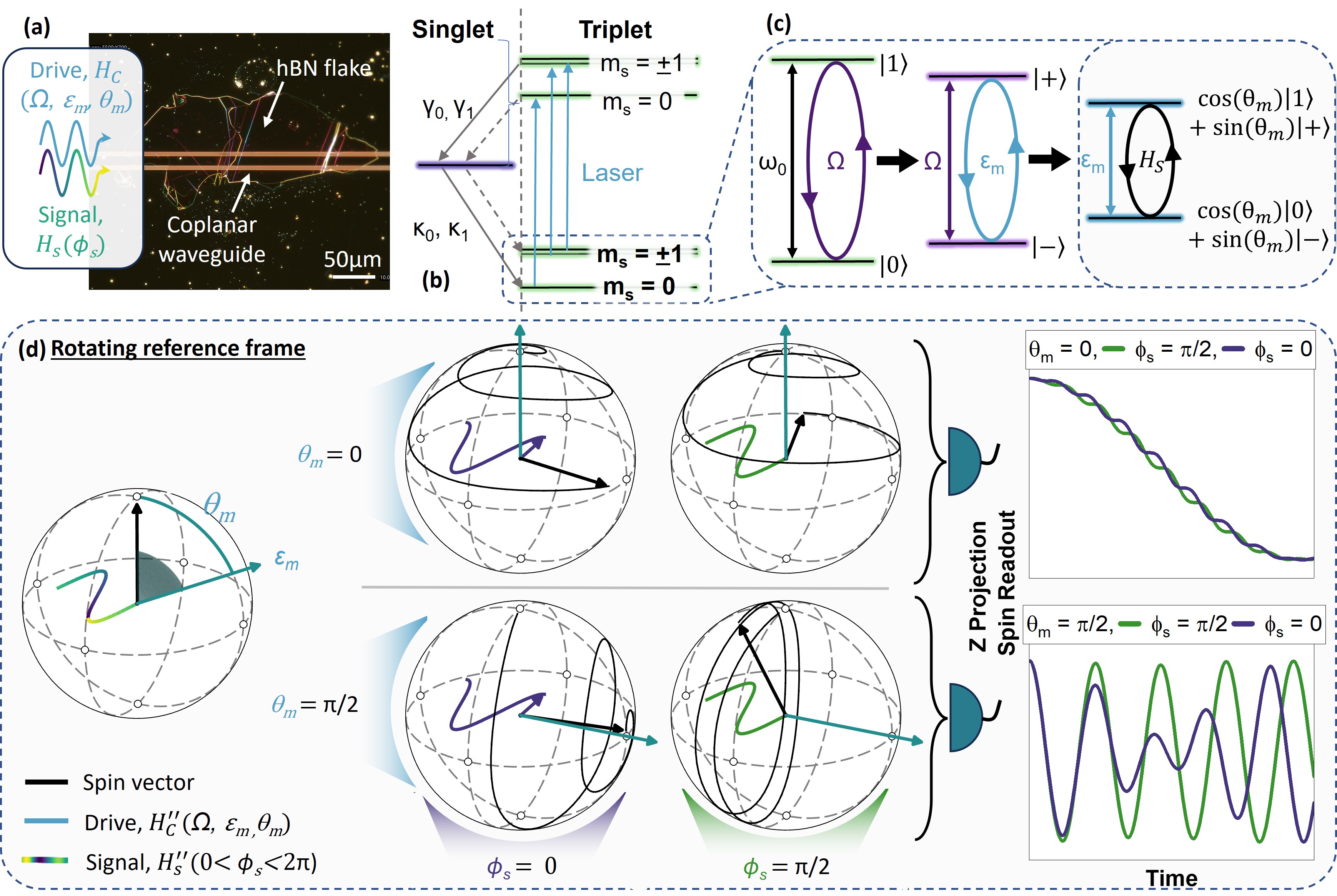}}
\caption{ \textbf{Overview of the sample and protocol.} \textbf{(a)} Darkfield microscope image of the device. The CCDD microwave drive, $H_c(\Omega, \epsilon_m, \theta_m)$ and signal field $H_s(\phi_s)$ are applied to the $V_B^{-}$ spin ensemble via the co-planar waveguide. \textbf{(b)} Simplified model of the $V_B^{-}$ optical and spin states, with spin dependent decay rates, $\gamma_0, \gamma_1, \kappa_0$ and $\kappa_1$. \textbf{(c)} The CCDD drive field resonantly addresses the two-level system corresponding to the $\vert m_s = 0\rangle$ to $\vert m_s = -1\rangle$ spin transition, producing concatenated dressed states defined by the drive amplitudes $\Omega$ and $\epsilon_m$, and drive phase $\theta_m$. A resonant signal $H_s$ will drive an additional transition between these states. \textbf{(d)} Signal driven $V_B^-$ spin evolution in a doubly rotating reference frame. The frame is selected so that the drive field, $H_c (\Omega, \epsilon_m, \theta_m)$ (turquoise arrow), reduces to a DC component in the YZ plane, with the polar angle determined by the phase of the drive, $\theta_m$. A resonant signal field applied along the x-axis exerts a torque on the spin vector (black arrow). The z-component of the subsequent spin trajectory is insensitive to signal phase if $\theta_m = 0$ (top), but is sensitive if $\theta_m = \frac{\pi}{2}$ (bottom). This can be detected via optical readout, which projects the spin onto the z-axis. To demonstrate this the spin trajectory is plotted under two different signal phases, $\phi_s = 0$ (blue) and $\pi/2$ (green).}
\label{fig1}
\end{figure*}

\section*{Methods}\label{sec:level2}
\subsection*{Experimental Setup}\label{sec:level3}
The device, shown in Fig. \ref{fig1}(a) is constructed of a sapphire substrate, patterned with a gold co-planar waveguide (CPW), which delivers the microwave drive fields to an ion-irradiated hBN flake on top. A 488nm laser is modulated by an acousto-optic modulator, and focused through an NA=0.55 microscope objective for optical excitation. The same objective collects photoluminescence, which is detected using a single photon avalanche detector (SPAD). The microwave drive and signal waveforms are generated using an arbitrary waveform generator (AWG). All results were recorded at room temperature in air. (see methods for further experimental details).

The boron vacancy hosts two unpaired electrons, producing a spin-1 triplet ground state aligned along the principal crystal axis. A simplified representation of this system and the relevant optical states are presented in Fig. \ref{fig1}(b). The $m_s = 0$ and $m_s =\pm 1$ spin states are separated by a zero field splitting (ZFS) of $D \approx 3.5 ~\mathrm{GHz}$, and a strain field of $E \approx 59 ~\mathrm{MHz}$\cite{Ramsay2023}. We use off resonant optical pumping to prepare the spin into the $m_s = 0$ state. The same optical excitation provides readout of the spin state, where the emitted photoluminescence (PL) is brighter for the $m_s = 0$ than the $m_s =\pm 1$ states. A DC-magnetic field applied along the c-axis of the hBN, $B_z\approx207 ~\mathrm{mT}$, produces a Zeeman shift of $\omega_{\pm 1} = \pm  \gamma_e B_z \approx \pm 5.74 ~\mathrm{GHz}$. The frequency of the $m_s=0 \Leftrightarrow m_s=-1$ transition used in these experiments is then $\omega_0 = D - E - \omega_{-1} = 2.32~\mathrm{GHz}$.

The hBN is of natural isotopic composition, with approximately $99.6\% ~\mathrm{^{14}N}$ nuclei, which has a nuclear spin $I=1$. The boron vacancy couples to the three nearest nitrogen nuclei\cite{Haykal2021}, with a strong hyperfine interaction (HFI) of $47 ~\mathrm{MHz}$. Fluctuations in the nuclear spin bath \cite{Gong2022} result in short $T_2^*$ coherence times, typically below $100 ~\mathrm{ns}$. We have shown previously that a strong CCDD microwave drive of $100 ~\mathrm{MHz}$ can mitigate the effects of this inhomogeneous noise and extend the coherence time to the few microsecond range\cite{Ramsay2023}.

\subsection*{Phase detection using a double microwave drive}\label{sec:level4}

The central idea of CCDD is to use a pair of microwave drives to isolate the spin from magnetic noise. The first drive protects against low frequency phase noise. If applied in quadrature, the second drive counteracts fluctuations in the first drive \cite{Cai2012}. When used in this way, a resonantly driven spin vector remains co-aligned with the second drive, creating a protected eigenstate. However, if exposed to a resonant signal field, effectively a third drive, the spin vector will deviate. The spin-projection along the z-axis is then detected optically and forms the basis of CCDD sensing schemes \cite{Stark2017, Wang2021, Patrickson2024}. 

\begin{figure*} 
\includegraphics[width=1\columnwidth]{{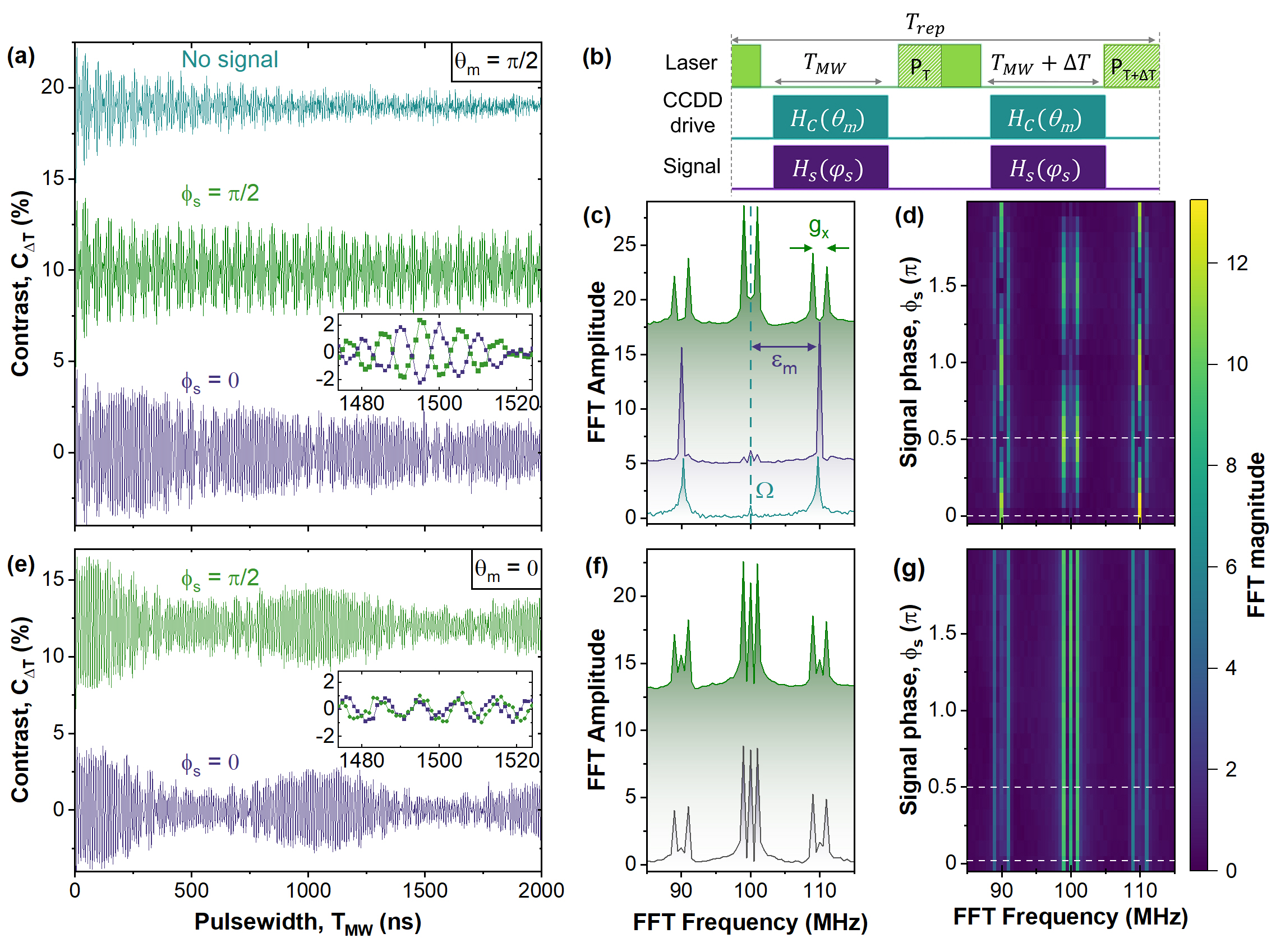}}
\caption{\textbf{$\bm{V_B^-}$ spin dynamics driven by a CCDD field, $\bm{H_c(\theta_m)}$, and a signal, $\bm{H_s(\phi_s)}$.} The signal is resonant with the $\omega_s = \omega_0 - \epsilon_m = 2.31 ~\mathrm{GHz}$ sensor transition. \textbf{(a)} \textbf{Phase-sensitive detection} Rabi measurements with the drive phase $\theta_m = \frac{\pi}{2}$, plotted without (turquoise) and with a signal applied for signal phases of $\phi_s = 0$ (blue) and $\phi_s = \frac{\pi}{2}$ (green). The inset highlights both the extended spin coherence in the presence of the signal, and the different sensor responses for the two signal phases. \textbf{(b)} Experimental sequence used to record the plots shown in (a) and (c)-(g). The contrast is recorded as $\mathrm{C_{\Delta T}} = \mathrm{(P_T - P_{T + \Delta T})/ P_{T + \Delta T}}$, where the reference readout is measured with $\Delta T = 5 \:\mathrm{ns} = \frac{\pi}{\omega_m} $ to cancel the effects of $T_1$ decay \cite{NanoLett2022_Baber}. \textbf{(c)} Fourier transforms of (a). The signal enhances the FFT amplitude for signal phases of $\phi_s = 0$, and creates additional Fourier components for $\phi_s = \frac{\pi}{2}$, where the spectrum is offset for clarity. \textbf{(d)} Equivalent of (c) for signal phases of $0 < \phi_s < 2\pi$, showing a smooth transition between two response regimes defined by $\phi_s = n\pi$ and $\phi_s = \frac{n\pi}{2}$. Dashed white lines correspond to spectra shown in (c). \textbf{(e)} \textbf{Phase insensitive detection} Rabi measurement where the drive phase $\theta_m = 0$. \textbf{(f)} Fourier transform of (e). \textbf{(g)} Fourier response for signal phases $0 < \phi_s < 2\pi$. Dashed white lines correspond to spectra shown in (f).}
\label{fig2} 
\end{figure*}

Whilst the signal phase influences the resulting spin trajectory, it has little impact on the z-projection (see Fig. \ref{fig1} (d)) and is therefore difficult to detect with optical readout techniques used in typical CCDD sensing schemes\cite{Stark2017, Wang2021, Patrickson2024}. To sense the signal phase, we apply the second drive in phase with the first drive, perpendicular to the spin vector, so that it modulates the Rabi frequency. This reduces the protection provided by the microwave drive \cite{Ramsay2023}. Instead, coherence stabilization is now provided by the signal field (Fig. \ref{fig2}(a)). The signal driven spin trajectory is again dependent on signal phase, but crucially, so is the projection along the z-axis (Fig. \ref{fig1} (d)).

The dynamics are determined by the system Hamiltonian. Choosing to resonantly drive the $m_s=0 \Leftrightarrow m_s=-1$ transition we approximate a two-level system, $H_0=\frac{1}{2}\omega_0\sigma_z$. The CCDD drive field, $H_{C}$, acts along the x-axis and in general, the signal field $H_s$ can be applied along an arbitrary axis,

\begin{eqnarray}
\begin{gathered}[b]
H = H_0 + H_{C} + H_s, \\
H_{C} =  \Omega\cos{(\omega_0 t - \frac{2\epsilon_m}{\Omega}\sin{(\omega_m t-\theta_m)}})\sigma_x, \nonumber \\  H_s = (g_x\sigma_x + g_y\sigma_y + g_z\sigma_z)\cos(\omega_s t + \phi_s)
\end{gathered}
\end{eqnarray}

where $\omega_0$ is the energy gap of the two level system, $\Omega$ and $\epsilon_m$ describe the amplitudes of the first and second CCDD drive fields, respectively.
The dynamics of the optically detected spin z-component are sensitive to the signal amplitude $g_i$, the signal frequency $\omega_s$ and, if the drive phase is $\theta_m = \frac{\pi}{2}$, the signal phase, $\phi_s$. 

We interpret this interaction by considering a doubly rotating reference frame, first with respect to $\frac{1}{2} \omega_0 \sigma_z$, and then $\frac{1}{2} \omega_m \sigma_x^\prime$ (primes denote the reference frame). This is a rotating frame that tracks the Rabi oscillation driven by $\Omega$. We focus on $g_x$, taking $g_y=g_z = 0$, as each signal vector drives similar spin dynamics but for different sensor resonances and signal phases (see Supplementary Note 3). Following the method in ref. \citenum{Patrickson2024}, for the case $\Omega = \omega_m$ we find,

\begin{eqnarray}
\begin{gathered}[b]
H^{''} = H_{C}^{''} + H_s^{''} = \frac{\epsilon_m}{2}[\sin{(\theta_m)}\sigma_y^{''}+\cos{(\theta_m)}\sigma_z^{''}] + \frac{g_x}{2}(\sigma_x^{\prime\prime}\cos((\omega_s - \omega_0)t + \phi_s))
\label{Eq:H''}
\end{gathered}
\end{eqnarray}

where we have applied the rotating wave approximation and assumed $\epsilon_m << \Omega$.
The CCDD drive reduces to a time-independent magnetic field, $\epsilon_m$, which points in the $YZ^{''}$ plane at a polar angle described by the phase of the drive, $\theta_m$. This is illustrated in Fig. \ref{fig1}(d). A benefit of using this approach is that the CCDD dressed spin states produce six tuneable transitions centred on the electron spin resonance. This means that our device can select signal frequencies across a $\sim300~\mathrm{MHz}$ range at a fixed DC field\cite{Patrickson2024} (also see Supplementary Note 2). In Eq. \ref{Eq:H''} we omit five of the resonances and consider only the $\omega_s = \omega_0 - \epsilon_m$ resonance, as the sensor couples more strongly to the signal field here\cite{Salhov2023, Patrickson2024}. For other sensor resonances, including off-resonant signals, see Supplementary Note 2.

$H^{''}$ sets the spin trajectory via the Heisenberg equation, $ \dot{ \sigma^{''}}=i[H^{''}, \sigma^{''}]$. To provide a more intuitive interpretation, we re-express this as the rotation $\vec{\dot{\sigma}}^{''} = \vec{H}^{''}(t) \times \vec{\sigma}^{''}$. To make the device sensitive to signal phase, we cast $\epsilon_m$ along the $y^{''}$-axis by choosing $\theta_m = \frac{\pi}{2}$. This causes the spin vector to rotate in the $XZ^{''}$ plane at the frequency $\epsilon_m$. Mixing the rotating spin with a signal oscillating at the frequency $\epsilon_m$ in the rotating frame will produce a constant torque on the spin vector, causing it to rotate. Crucially, the coupling strength depends on the signal phase. To illustrate this interaction, in Fig. \ref{fig1}(d) we plot the trajectories of two different spin vectors exposed to signals of phase $\phi_s = 0$ (blue), and $\phi_s = \frac{\pi}{2}$ (green) (see methods). Evolution occurs in the same doubly rotating frame described above. The torque produced by the signal is largest for $\phi_s = 0$, causing a rapid divergence of the spin vector from its original trajectory, which lay in the $XZ^{''}$ plane. The coupling strength is minimised for $\phi_s = \frac{\pi}{2}$. The cumulative impact on $\dot{\sigma}$ can be seen after $\epsilon_m t = 4\pi$, where the z-projection of the spin vector is drastically different for the two cases. 

\section*{Results}
\subsection*{Phase sensitive microwave detection}

\begin{figure*} 
\includegraphics[width=1\columnwidth]{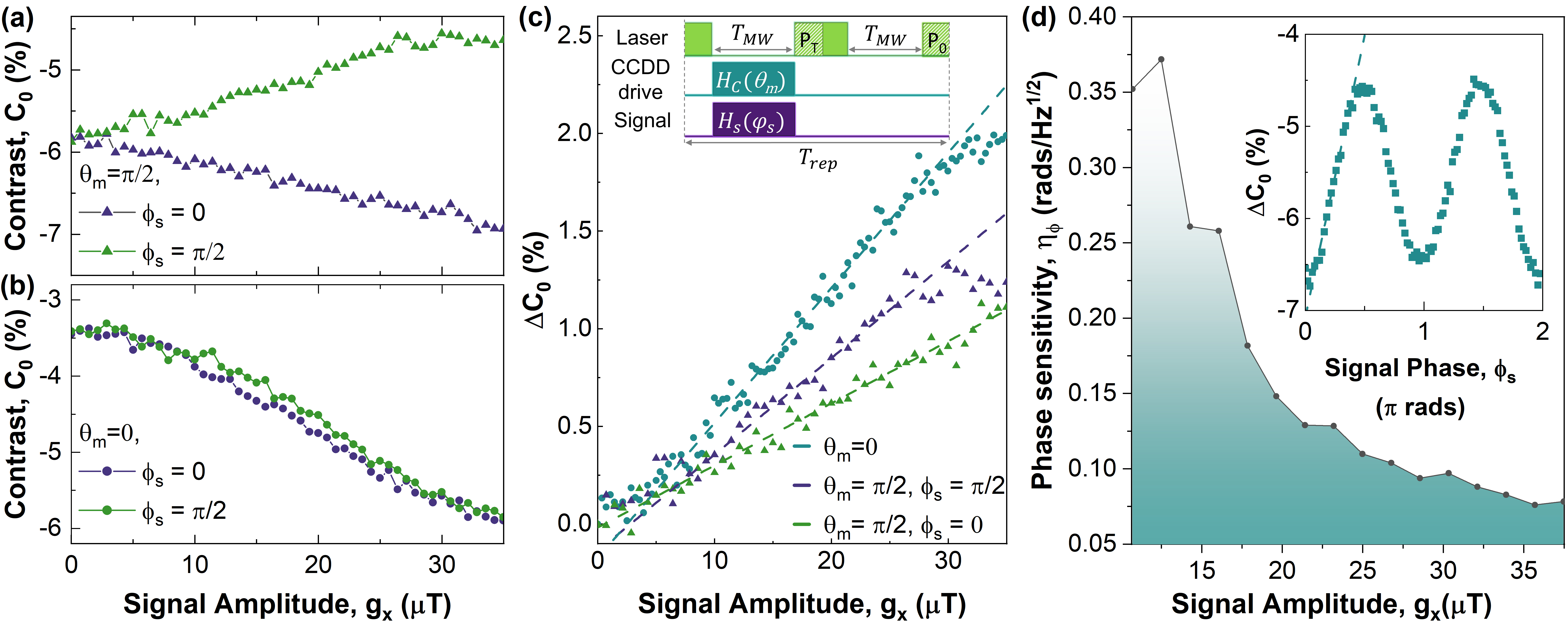}
\caption{\textbf{Benchmarking of the sensor response to signal amplitude and phase.} Contrast as a function of signal amplitude for $\theta_m = \frac{\pi}{2}$ in \textbf{(a)} and $\theta_m = 0$ in \textbf{(b)}, plotted for signal phases of $\phi_s = 0$ (blue), $\frac{\pi}{2}$ (green). The inset of (c) describes the experimental sequence used for (a)-(d), with a fixed pulsewidth of $T_{MW} = 950 \mathrm{ns}$, and the contrast calculated as $\mathrm{C_0} = \mathrm{(P_T - P_0)/ P_0}$. \textbf{(c)} The change in contrast with and without a signal applied as a function of signal amplitude. Linear fits provide ${\max\lvert\frac{\partial (\Delta C)}{\partial B}\rvert}$ for the sensitivity calculation in Eq. \ref{Eq:Sensitivity}. \textbf{(d)} Phase sensitivity as a function of signal amplitude. A value for ${\max\lvert\frac{\partial (\Delta C)}{\partial \phi_s}\rvert}$ was required for each phase sensitivity calculation. The measurement used for a signal amplitude of 0.9 MHz is provided in the inset as an example.}
\label{fig3}
\end{figure*}

To probe the CCDD driven spin dynamics in the presence of a resonant signal, we perform Rabi experiments with the pulse sequence shown in Fig. \ref{fig2}(b). The sequence samples the z-component of the spin trajectory after simultaneously applying a control and signal microwave pulse of duration, $T_{MW}$ ($T_{MW} + \Delta T$), with optical readout measurements $\mathrm{P_T}$ ($\mathrm{P_{T + \Delta T}}$). These values are then used to calculate a normalised contrast, $\mathrm{C_{\Delta T}} = \mathrm{(P_T - P_{T + \Delta T})/ P_{T + \Delta T}}$, where $\Delta T = 5 \mathrm{ns} = \frac{\pi}{\omega_m} $ to nullify the effects of $T_1$ decay. In Fig. \ref{fig2}(a) we use the sequence to illustrate the spin response when the drive phase $\theta_m = \frac{\pi}{2}$. Note that without a signal (turquoise) the ensemble decoheres within $\sim 1000$ ns due to unprotected fluctuations in the CCDD drive amplitude $\Omega$ \cite{Ramsay2023}. However, in our scheme this is corrected by the signal field, which behaves as an additional decoupling drive. This improves the devices sensitivity by extending spin coherence. This is shown in Fig. \ref{fig2}(a) for a signal amplitude of $g_x$ = 2 MHz, frequency $\omega_s = \omega_0 - \epsilon_m$ = 2.31 GHz, and signal phases of $\phi_s = 0$ (blue) and $\phi_s = \frac{\pi}{2}$ (green). The inset depicts the marked difference in spin response for the two phases, and can be visualised by the Fourier transforms in Fig. \ref{fig2}(c) where we find a pair of nested Mollow triplets. The central frequency is produced by the CCDD drive at $\Omega = 100 \,\mathrm{MHz}$. The sidebands at $\Omega \pm \epsilon_m =$ 90 and 110 MHz are produced by mixing with the second CCDD drive. The splitting of the central peak and sidebands is caused by the signal field $g_x/2 = 1 \,\mathrm{MHz}$ \cite{Patrickson2024}. These dynamics reflect the trajectories plotted in Fig. \ref{fig1}(d), however they are superimposed onto the frequency components of the rotating reference frame, introducing additional complexity in the spin response. Note that the Fourier frequencies are independent of (dependent on) signal amplitude, $g_x$, for signal phases of $\phi_s = n\pi$ ($\phi_s = \frac{n\pi}{2}$), where n is an integer (see Supplementary Note 4). To highlight the distinction with CCDD sensing schemes that omit drive phase, in Fig. \ref{fig2}(e) we plot Rabi measurements with the drive phase $\theta_m = 0$. Here the plots are qualitatively the same, and the Fourier transforms in Fig. \ref{fig2}(f) and (g) confirm that the sensor response is independent of signal phase.

To operate the device as a sensor we use the pulse sequence depicted in Fig. \ref{fig3}(b). The contrast $\mathrm{C_0} = \mathrm{(P_T - P_{0})/ P_{0}}$ is recorded at a single pulsewidth of $T_{MW} = 950~\mathrm{ns}$, which corresponds to a peak in the CCDD Rabi oscillation \cite{Degen2017}. The reference readout $P_0$ is collected with the CCDD MW drive, $H_c(\theta_m) = 0$, and signal, $H_s(\phi_s) = 0$. Turning the CCDD drive off is useful in the case of an unknown continuous signal as the dressed spin states are removed, providing a robust reference measurement that is unresponsive to the signal field. In Fig. \ref{fig3}(a) we record the contrast $\mathrm{C_0}$ as a function of signal amplitude for signal phases of $\phi_s = 0, \frac{\pi}{2}$. The diverging responses demonstrates the sensitivity to signal amplitude and phase. In Fig. \ref{fig3}(b) we again compare to the case of $\theta_m = 0$, where the response is sensitive to signal amplitude, but not signal phase. 

To benchmark the sensors performance, we calculate the amplitude sensitivity, $\eta$,

\begin{eqnarray}
\eta = \frac{\mathrm{S}(t_{m})}{\max\lvert\frac{\partial (\Delta C_0)}{\partial g_x}\rvert} \sqrt{t_m}
\label{Eq:Sensitivity}
\end{eqnarray}

where $g_x$ is the signal amplitude, $\Delta C_0 = \vert(C_0(g_x) - C_0 (g_x = 0)\vert$ describes the change in contrast due to the signal field, S$(t_{m})$ is the standard deviation in $\Delta C_0$, and $t_{m}$ is the measurement time. To calculate ${\max\lvert\frac{\partial (\Delta C_0)}{\partial g}\rvert}$ we plot the change in contrast, $\Delta C_0$ as a function of signal amplitude, $g_x$, for different drive and signal phases in Fig. \ref{fig3}(c). Each data point is averaged over 10 measurements to provide an estimate of the standard deviation, $S(t_{m})$. 

For $\theta_m = \frac{\pi}{2}$, this gives sensitivities of $5.1~\mathrm{\mu T/ \sqrt{Hz}}$ and $3.4~\mathrm{\mu T/ \sqrt{Hz}}$ for signal phases of $\phi_s = 0$ and $\phi_s = \frac{\pi}{2}$, respectively. These values are comparable to other AC magnetometry schemes using $V_B^-$ ensembles \cite{Rizzato2023, Patrickson2024}, and as a reference for our device we find $\eta = 2.5 \mathrm{\mu T/ \sqrt{Hz}}$ in the case of zero drive phase, $\theta_m = 0$. This suggests that the protocol is not only able to resolve signal phase, but does so with only a modest decrease in amplitude sensitivity.

The inset in Fig. \ref{fig3}(d) shows an analogous measurement for calculating the phase sensitivity $\eta_{\phi}$, with the drive phase $\theta_m = \frac{\pi}{2}$. For a fixed value of $T_{MW}$ the contrast $\Delta C_0$ has a strong dependence on signal amplitude, as it effects both the spin coherence and the Fourier components of the spin response. To characterise this relationship, we plot the phase sensitivity, $\eta_{\phi}$, as a function of signal amplitude, $g_x$, in the main panel of Fig. \ref{fig3}(d). A minimum in the phase sensitivity of $\eta_{\phi} = 0.076 \mathrm{rads/\sqrt{Hz}}$ is reached for signal amplitudes of $g_x \approx 35 \mathrm{\mu T}$. At the cost of amplitude sensitivity, this minimum could be improved by optimising $T_{MW}$ to reduce the total measurement time, $t_m$ (see Supplementary Note 5). Note that the signal amplitude can also be measured independent of signal phase by choosing $\theta_m = 0$\cite{Patrickson2024}. 

\begin{figure*}
\includegraphics[width=1\columnwidth]{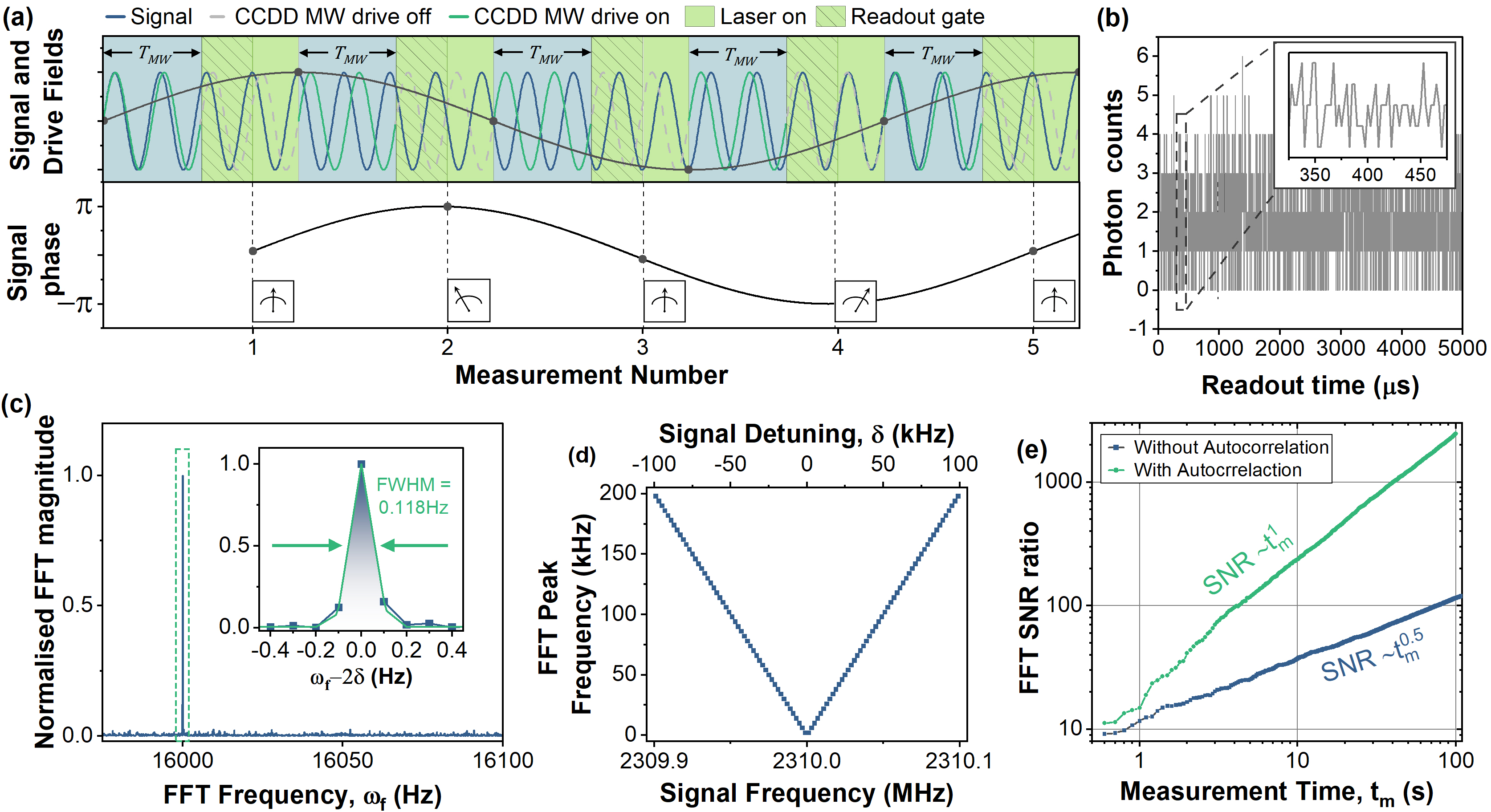}
\caption{\textbf{Quantum heterodyne measurements using a continuous microwave drive.} \textbf{(a)} Schematic of the experimental sequence. A coherent microwave drive (turquoise) is used across sequential measurements, each lasting $T_{MW}$, to record the phase of a continuous, coherent signal (blue). The sensor output (bottom panel) records the relative detuning. \textbf{(b)} Experimental photon time trace, giving an excerpt of the sensor output analogous to the bottom panel of (a). The total measurement time $t_m$ was 10 seconds, with an average of 1.8 photons collected per measurement. \textbf{(c)} FFT of the autocorrelation taken from (b). The inset shows a Gaussian fit providing a SNR of 235, with the FWHM giving a frequency resolution of 0.118 Hz. \textbf{(d)} FFT SNR as a function of total measurement time, $t_m$. Taking the FFT of the autocorrelated data improves the SNR scaling from $\sqrt{t_m}$ to $t_m$.}
\label{fig4}
\end{figure*}

\subsection{CCDD heterodyne sensing}

In the measurements of  Fig. 3 the sequence repetition time, $T_{rep} = 5~\mathrm{\mu s}$, and the phase measurement is averaged over $10^5$ measurements (total measurement time, $t_m=1~\mathrm{s}$). The signal phase is set to be the same at the beginning of each measurement period, as would be the case for an interferometric measurement of a signal exactly resonant with $\omega_0+\epsilon_m$. However, the phase sensitivity also allows this method to be adapted for quantum heterodyne sensing \cite{Schmitt_Science_2017, Boss_Science_2017}, where the instantaneous phase information is recorded for each individual measurement. Fig. \ref{fig4}(a) provides an overview of the CCDD quantum heterodyne sensing technique. The CCDD drive field acts as the local clock and provides a stable coherent phase reference, and is switched on for a time $T_{MW}$ during each sensing sequence. The instantaneous phase difference between the continuously applied signal and the local clock provided by the CCDD drive is encoded in the PL intensity originating from the subsequent laser pulse. In this way, the relative phase is tracked as a function of time, thereby sampling the beat frequency between signal and clock. Taking the Fourier transform reveals the detuning of the signal from the clock, with a frequency resolution determined by the total number of measurements multiplied by $T_{rep}$.

Fig. \ref{fig4}(b) shows the experimental realisation of the CCDD quantum heterodyne method, where an excerpt of a spin dependent PL time trace is plotted. To record the data, a continuous signal of frequency $2310.008~\mathrm{MHz}$ was generated by the AWG and applied via the CPW. The $\omega_0 + \epsilon_m =  2310~\mathrm{MHz}$ resonance provided the $H_c$ clock frequency. An average of 1.8 photons were recorded per readout sequence. The time sequence data is recorded for a total measurement time, $t_m = 10~\mathrm{s}$. The data is autocorrelated, then fast Fourier transformed \cite{Meinel2021}, as plotted in Fig. \ref{fig4}(c). The dominant Fourier component $\omega_f$ corresponds to the signal demodulated by the clock at $16 ~\mathrm{kHz}$, and has a full-width half maximum of 0.118 Hz. Note that the peak appears at twice the true detuned frequency of $\delta=\omega_s - \omega_0 - \epsilon_m = 8 ~\mathrm{kHz}$. This is because the sensor response cycles twice for signal phases between $0 \geq \phi_s \geq 2 \pi$, as can be seen in the inset of Fig. \ref{fig3}(d). Comparing the peak height to the standard deviation of the baseline gives a signal to noise ratio, SNR=235.

For a fixed set of CCDD drive parameters, the device can detect detuned signals within a 100 kHz range of the sensor resonance. This is demonstrated in Fig. \ref{fig4}(d), where we plot the peak frequency from the autocorrelation FFT as a function of signal detuning. This range is determined by the measurement repetition rate, $1/T_{rep} = 400 ~\mathrm{kHz}$, which sets the Nyquist frequency of the PL time trace. This limit could be extended by reducing $T_{MW}$ from $950 ~\mathrm{ns}$ to 10s of $\mathrm{ns}$ (see Supplementary Note 5) or by shifting the sensor resonance with the $H_c$ drive amplitude, $\epsilon_m$ \cite{Patrickson2024}. Although \ref{fig4}(c) shows that the sign of the detuning is not implicit from a single measurement, this can also be deduced using $\epsilon_m$. Finally, in Fig. \ref{fig4}(d) we characterise how the SNR scales with measurement time, confirming that $\mathrm{SNR} \propto t_m$ (green) with autocorrelation, compared to a shot noise limited $\mathrm{SNR} \propto \sqrt{t_m}$ without (blue)\cite{Meinel2021}.

\section*{Discussion}\label{sec:level5}

To conclude, we have demonstrated a phase modulated CCDD sensing protocol that extends the coherence time of a $V_B^-$ ensemble into the $\mathrm{\mu s}$ regime by suppressing the effect of magnetic noise from the nuclear spin bath, whilst providing tuneable control over the phase, frequency and amplitude response of the sensor. Benchmarking our sensor, we measure amplitude and phase sensitivities of $\eta \approx 3-5 \:\mathrm{\mu T \sqrt{Hz}}$ and $\eta_{\phi} \approx 0.076 \:\mathrm{rads \sqrt{Hz}}$, respectively. Finally, the sensing protocol is embedded in a quantum heterodyne measurement, where we measure a frequency resolution below 1 Hz for a signal frequency of $\sim$2.31 GHz. 

Our quantum heterodyne protocol employs a continuous concatenated dynamic decoupling sequence, in contrast to previous implementations using pulsed sequences\cite{Boss_Science_2017, Schmitt_Science_2017, Glenn2018, Jiang2023}. The majority of these schemes are limited to signal frequencies below $\sim 50 ~\mathrm{MHz}$ because they require the electron spin flip rate to be comparable to the signal frequency. This requires complex, high power pulse sequences susceptible to accumulated pulse-area errors which can impair the sensitivity\cite{Ishikawa2018, Cao2020, Lang2019, Aharon2019}. The well established CASR protocol\cite{Glenn2018} was recently demonstrated using a $V_B^-$ ensemble in hBN for the first time\cite{Rizzato2023}, where the authors achieved a frequency resolution of 0.9 Hz for an 18 MHz signal sampled for 2000 $\mathrm{s}$, compared with 0.118 Hz resolution for a 10 s integration in this work. Note that, whilst the results presented here focused on GHz signals, the same protocol and experimental setup is also capable of detecting signals in the 10-150 MHz range \cite{Patrickson2024}.

Newer pulsed protocols capable of detecting GHz signals have been proposed\cite{Chu2021} and demonstrated\cite{Meinel2021} using NV centres in diamond, however these still suffer from the aforementioned complexity and power constraints. A continuous scheme is also presented in \cite{Meinel2021}, however, the single drive does not extend the coherence time, limiting the sensitivity. Other continuous heterodyne schemes take advantage of the natural $\mathrm{\mu s}$ coherence times inherent to NV centres in isotopically purified diamond\cite{Staudenmaier2021, Wang2022, Li2023}. However these are not an option with $V_B^-$ ensembles, where the natural spin state lifetime is $<100 \:\mathrm{ns}$ \cite{NanoLett2022_Baber, Haykal2021}. In contrast, our scheme is able to extend $V_B^-$ coherence times into the $\mathrm{\mu s}$ regime, whilst remaining sensitive to signal phase, frequency and amplitude. It is important to acknowledge the significance of achieving this in a two-dimensional material, which could enable nanoscale sensor-source distances unaffected by dangling bonds and surface imperfections in the host material.

This system can be applied as an effective probe of other low dimensional condensed matter systems that present AC magnetic fields\cite{Bertelli2020, Lee-Wong2020}. For instance, the out of plane DC field used in this work would support forward volume spin wave modes in ferromagnetic thin films\cite{Chumak2015, Chen2019}. These spin waves are a promising platform for next generation data transfer\cite{Chumak2015}, as they avoid ohmic heat loss and support GHz to THz fields. For example, spin waves resonant with our device could be optically excited\cite{Hashimoto2017, Au2013} at one end of the CPW. As these spin waves propagate with $\le 100 ~\mathrm{\mu m}$ wavelengths\cite{10.1063/1.5019752}, the phase dependent sensor response could be used with time correlated laser scanning confocal microscopy to image the spin waveform along the length of the CPW, potentially providing new insight into spin wave dispersion relations\cite{Divinskiy2021, Hashimoto2017}. This sensing scheme could also aid in the development of microwave circuitry, to probe failure modes or ohmic heating in regions of high current loads\cite{Bohi2010, Garsi2024, Nowodzinski2015}. More broadly, combining CCDD schemes with bright single spin defects\cite{Stern2022}, and using local nuclear spins as ancilla qubits, presents a promising platform for future quantum sensing endeavours.

\backmatter

\section{{Methods}}

\subsection{Experimental}
PL is excited using a 488 nm diode laser pulsed by an acousto-optic modulator. An objective lens (N.A.=0.55) is used to focus the laser to a diffraction-limited spot on the hBN flake and at the center of the co-planar waveguide. Photoluminescence from the boron vacancy ensemble is collected with the same objective, filtered by a 750 nm long pass filter and recorded on a single photon avalanche diode (SPAD). The microwave control and signal waveforms are generated using an arbitrary waveform generator, amplified and applied via a circulator to the CPW. The other end is terminated 50 ohms. The optical and microwave excitation are synchronised with a digital pattern generator and the photon counts recorded with time-tagging electronics. All experiments applied a constant phase offset of 0.07$\pi$ to the drive field, as this produced the optimum sensor response to signal phase in the device.

The maximum waveform length is limited by the AWG memory to approximately 2.5 ms. For simplicity, we choose a waveform length of 1 ms and ensure the frequency of all control and signal components is an integer multiple of 1 kHz. The 1 ms sequence is repeated resulting in signals with greater coherence than dictated by the memory constraints [\cite{Meinel2021}]. The optical and microwave control pulses are applied to the sample, along with the constant signal field for a total measurement time, $t_m$. All detected photons are time-tagged and saved. In post-processing a time gate is applied to keep only photons that arrive within a time $t_{gate}=~\mathrm{350 ns}$ of the beginning of each optical pulse (repetition rate of 400 kHz). These photons are then binned according to the time that has elapsed since the beginning of the measurement, with an example shown in Fig. \ref{fig4}(b). We fast Fourier transform the autocorrelation of this data to generate a spectrum of the signal, down-converted by the frequency of the double-dressed resonance, $\omega_0-\epsilon_m$.  

\subsection{Model}

The Bloch-vector dynamics plotted in Fig. \ref{fig1}(d) were numerically calculated using the Heisenberg equation $\dot{\sigma^{'}}=i[H_C^{'},\sigma^{'}]$, which is re-expressed as a rotation, $\vec{\dot{\sigma}} = \vec{H}^{''}(t) \times \vec{\sigma}$, where $\vec{\dot{\sigma}}$ is the time derivative of the spin vector, $\vec{H}^{''}(t)$ is the interaction picture Hamiltonian as defined in the main text, and $\vec{\sigma}$ is the spin vector. For each time step used in the model, a single effective field is calculated by first summing the vector components in $\vec{H}^{''}(t)$. This forms an axis rotation for the spin vector, with the angle of rotation determined by the magnitude of the summed vector components. The process is repeated iteratively for all time steps to produce the plotted trajectories.

\bmhead{Acknowledgments}
This work was supported by the Engineering and Physical Sciences Research Council [Grant numbers EP/S001557/1 and EP/L015331/1], Partnership Resource Funding from the Quantum Computing and Simulation Hub [EP/T001062/1] and an Engineering and Physical Sciences Research Council iCASE in partnership with Oxford Instruments Plasma Technology. Ion implantation was performed by Keith Heasman and Julian Fletcher at the University of Surrey Ion Beam Centre. For the purpose of open access, the author has applied a ‘Creative Commons Attribution (CC BY) licence to any Author Accepted Manuscript version arising’.

\begin{appendices}




\end{appendices}


\providecommand{\noopsort}[1]{}\providecommand{\singleletter}[1]{#1}%

\end{document}


\maketitle

\subsection{\label{sec:1}{$V_B^-$ spin characterisation}}

\begin{figure*}[h!] 
\centering
\includegraphics[width=0.5\columnwidth]{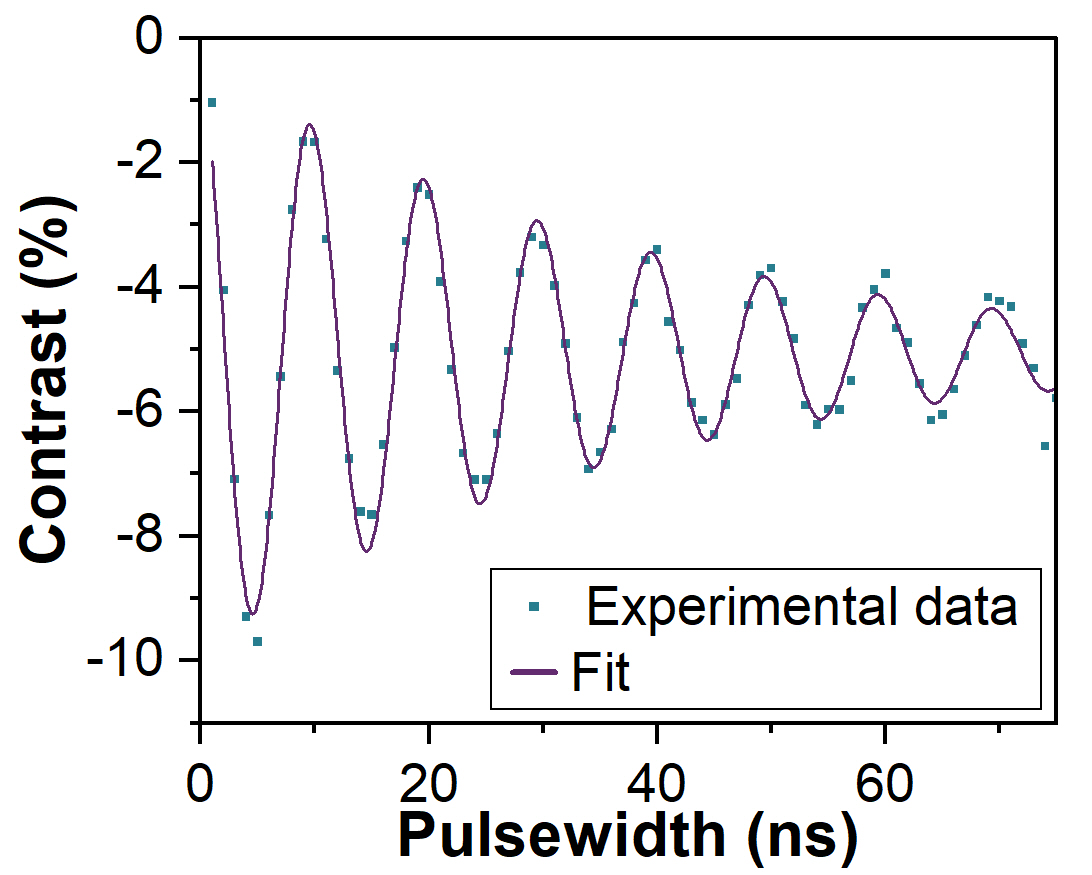}
\caption{\textbf{$V_B^-$ Rabi measurement} Rabi oscillation of the $m_S = 0$ to $m_S = -1$ ground state transition. A fit to a damped sine $y = y_0 + \sin(\omega t)e^{-t/ T_{Rabi}}$ gives a coherence time of $T_{Rabi}$ = 36ns.}
\label{SIfig1}
\end{figure*}


\subsection{\label{sec:2}{Device Response to Signal Frequency}}

\begin{figure*} 
\centering
\includegraphics[width=0.5\columnwidth]{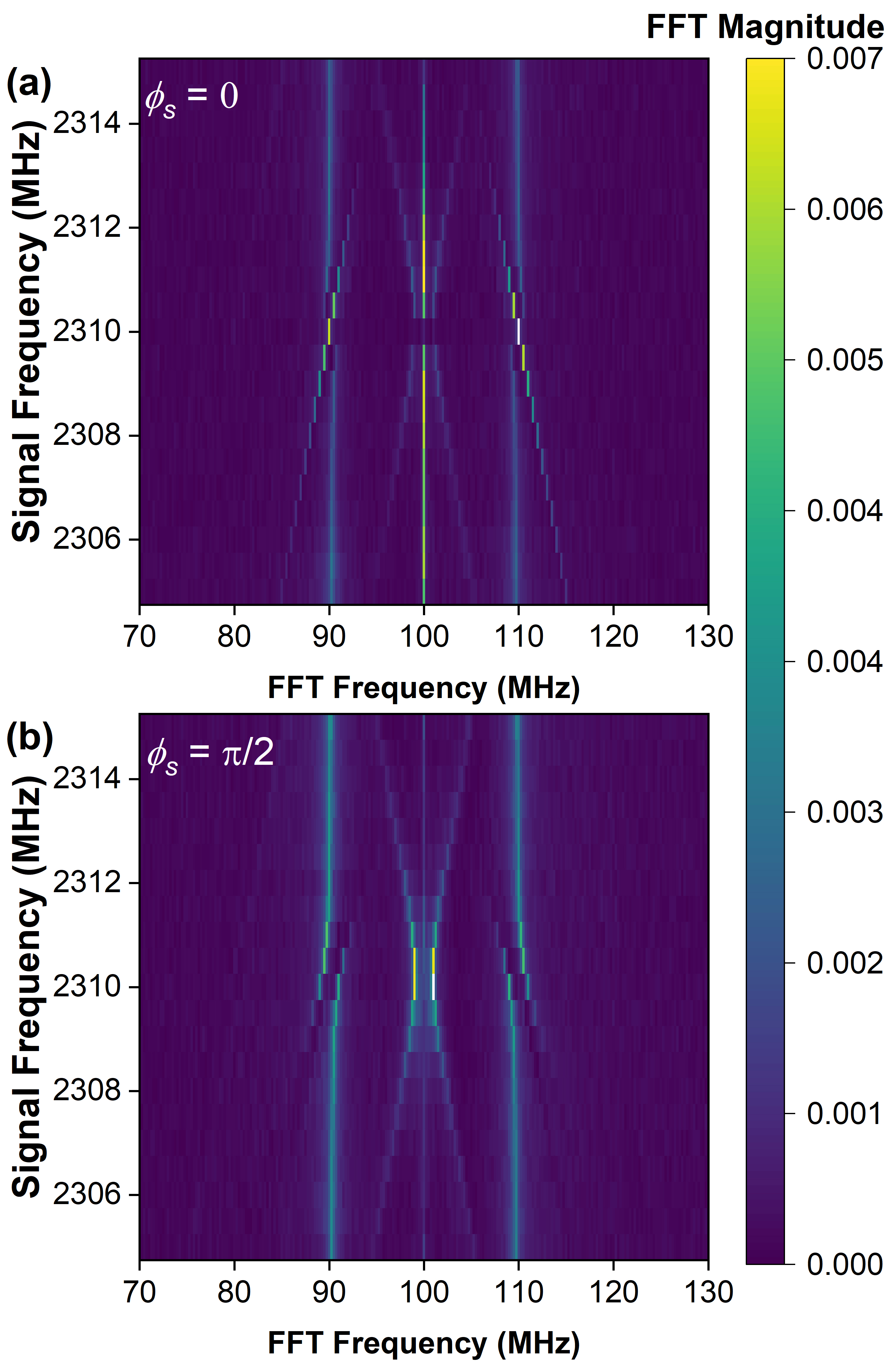}
\caption{\textbf{Off resonant sensor response}. Fourier transforms of CCDD driven Rabi oscillations as a function of applied signal frequency for signal phases of $\phi_s = 0$ in \textbf{(a)} and $\phi_s = \frac{\pi}{2}$ in \textbf{(b)}. The sensor resonance is set to $\omega_s = \omega_0 - \epsilon_m = 2.31 \:\mathrm{GHz}$. When this condition is met a sharp change in the Fourier response is seen for both signal phases, indicating the device sensitivity to signal frequency. Further information can be found at \cite{Patrickson2024}}
\label{SIfig2}
\end{figure*}

\begin{figure*}[h!] 
\centering
\includegraphics[width=0.5\columnwidth]{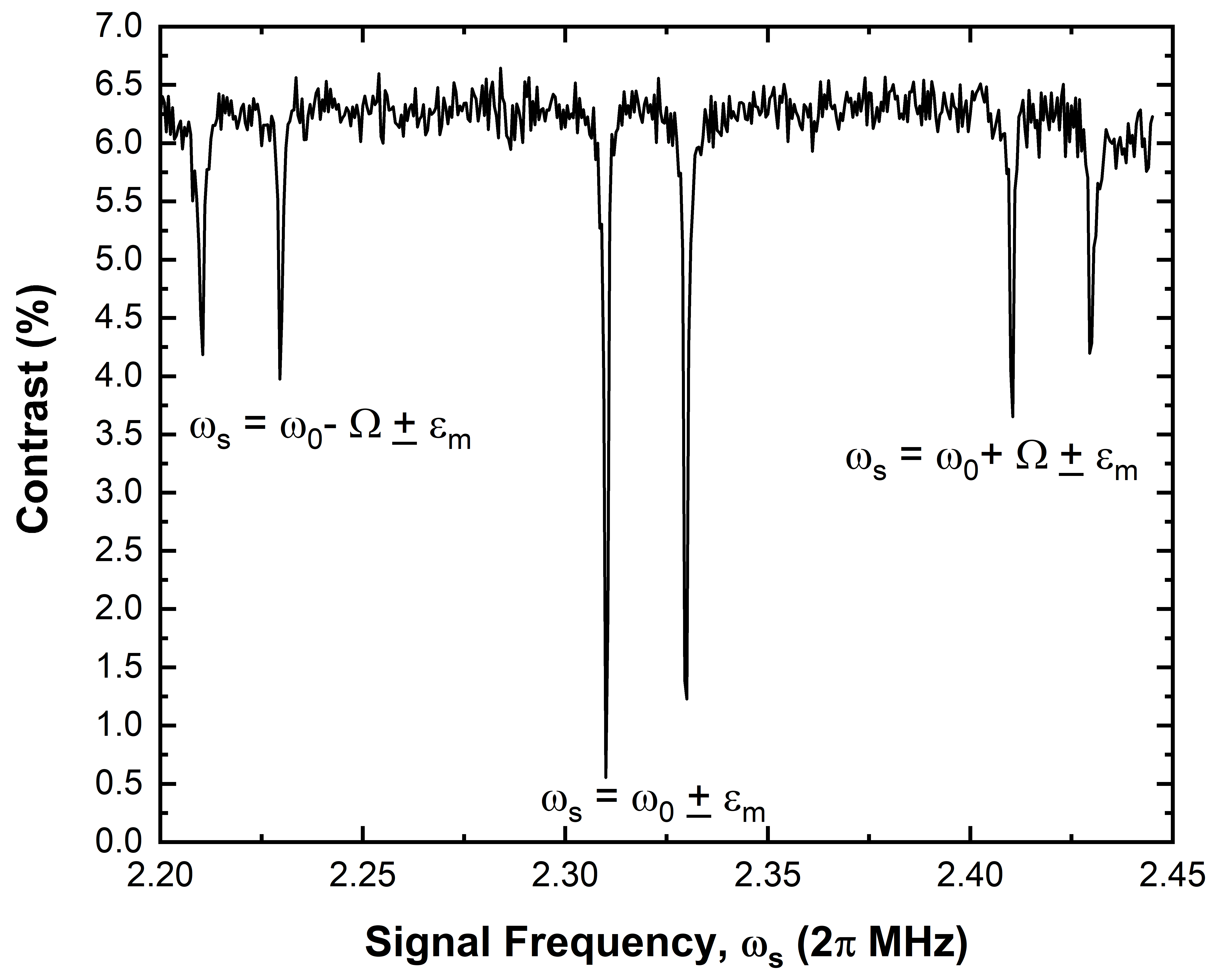}
\caption{\textbf{Sensor resonances} Sensor response as a function of signal frequency $\omega_s$, displaying six sensor resonances centered on the electron spin transition, $\omega_0 = 2.32$ GHz. The contrast was sampled after exposing the $V_B^-$ ensemble to a CCDD pulsewidth of $T_MW = 950$ ns. The applied signal had an amplitude of $g_x = 0.8$ MHz and phase $\phi_s = 0$. Drive phase was omitted, $\theta_m = 0$.}
\label{SIfig0}
\end{figure*}

In this section we start by discussing the sensor response to off-resonant signals. We perform CCDD Rabi oscillations analogous to Fig. 2 of the main text, but as a function of signal frequency. The interaction is illustrated by the FFTs plotted in Supplementary Fig. \ref{SIfig2}(a) and (b) for signal phases of $\phi_s = 0$ and $\frac{\pi}{2}$, respectively. The Fourier response centres on the same nested Mollow triplet structure as in the main text, with a central frequency of $\Omega = 100 \:\mathrm{MHz}$, CCDD sidebands at $\Omega \pm \epsilon_m = 100 \pm 10 \:\mathrm{MHz}$ and signal induced sidebands. The latter is complex in structure, and depends on the detuning between the sensor resonance and the signal frequency, $\delta = \omega_s - \Omega_0 - \epsilon_m$. The sensor undergoes a sharp transition when exposed to a resonant signal, shown here at a signal frequency of 2.31 GHz, with distinct responses for each signal phase. In particular, the main frequency components change from $\Omega = 100 \:\mathrm{MHz}$ to $\Omega \pm \epsilon_m = 100 \pm 10 \:\mathrm{MHz}$ for signal phases of $\phi_s = 0$, and from $\Omega \pm \epsilon_m = 100 \pm 10 \:\mathrm{MHz}$ to $\Omega \pm g_x = 100 \pm 1 \:\mathrm{MHz}$ for signal phases of $\phi_s = \frac{\pi}{2}$. This illustrates our protocols dependence on signal frequency, which we detect by effectively filtering between these different Fourier regimes. Far from resonance, the response reduces to frequencies at $\Omega \pm \epsilon_m$, as expected for a CCDD Mollow triplet with the drive phase $\theta_m = \frac{\pi}{2}$ \cite{Ramsay2023}.

\begin{figure*} 
\centering
\includegraphics[width=1\columnwidth]{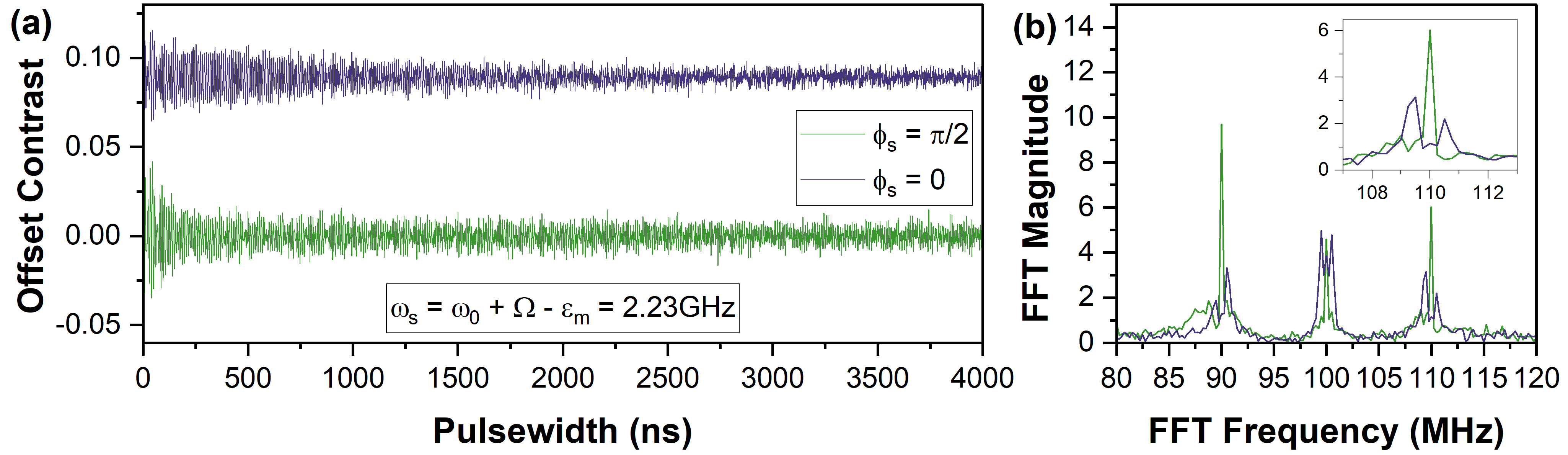}
\caption{\textbf{Alternative sensor resonance.} \textbf{(a)} CCDD driven Rabi oscillation with the same parameters as in the main text, but for a signal frequency of $\omega_s = 2.23\: \mathrm{GHz}$ instead of $\omega_s = 2.31\: \mathrm{GHz}$. The two resonances display analogous behaviour, demonstrating that the additional tuneable resonances produced through the CCDD scheme are also sensitive to signal phase. \textbf{(b)} Fourier transform of (a). The inset shows a closeup of the right hand Mollow triplet, which presents an opposite phase response to the resonance used in the main text.}
\label{SIfig3}
\end{figure*}

The CCDD microwave drive also produces multiple other sensor resonances. Each can be tuned using the drive parameters and display a similar response \cite{Patrickson2024}. This is an important feature of CCDD sensing schemes, as tuning the electron spin transition normally involves changing the DC magnetic field by moving an external magnet, which is slow and imprecise. Whereas in our device, the sensor resonances can be tuned within $\pm$150 MHz of the electron spin transition electronically, as illustrated in Supplementary Fig. \ref{SIfig0}. This is useful when probing signals of unknown frequency. 

Using our protocol, in Supplementary Fig. \ref{SIfig3} we show that these additional resonances can also resolve signal phase. We use the same CCDD parameters as those in the main text, which targeted a resonance at $\omega_s = \omega_0 - \epsilon_m = 2.31 \:\mathrm{GHz}$. Instead, here we apply a signal frequency of $\omega_s = \omega_0 - + \Omega \epsilon_m = 2.23 \:\mathrm{GHz}$ and analyse the response of a CCDD Rabi measurement. To illustrate the dependence on signal phase, Supplementary Fig. \ref{SIfig3}(a) shows the response for $\phi_s = 0$ and $\frac{\pi}{2}$. The Fourier response is presented in Supplementary Figs. \ref{SIfig3}b), where its clear that each signal phase produces a different sensor output, analogous to the results presented in the main text. Note that for the two resonances the Fourier regimes are out of phase by $\frac{\pi}{2}$. This has no impact on the sensors ability to distinguish signal phase however, as the protocol only requires two distinct responses to contrast against. We also note that the magnitude of the Fourier response is reduced for signal frequencies of $\omega_s = \omega_0 - + \Omega \epsilon_m = 2.23 \:\mathrm{GHz}$. This applies for all resonances other than  $\omega_s = \omega_0 \pm \epsilon_m$, as they are subject to increased attenuation in the dressed state frame of reference \cite{Salhov2023}.

\subsection{\label{sec:3}{Spin Response to Signal Vector}}

CCDD sensing schemes operate by driving the spin vector along multiple axes at multiple frequencies, producing additional resonances in the system whilst also decoupling the spin from sources of magnetic noise. These dynamics produce a spin response that depends simultaneously on the signal frequency and it's direction of propagation. Our sensor displays six resonances centred on the electron spin resonance. The frequencies depend on the CCDD drive amplitudes $\Omega$ and $\epsilon_m$ and are only sensitive to signals propagating in the XY plane\cite{Patrickson2024}. Two additional resonances appear in the MHz range, which are sensitive to signals propagating along the z-axis, and depend only on $\Omega$ and $\epsilon_m$. In the main text we focused on a single resonance for a signal propagating along the x-axis. Here we consider three signals, one propagating along each axis. Selecting a single resonance for each we show that the protocol retains phase sensitivity regardless of the signal direction.

We consider the signal field,

\begin{equation}
\begin{gathered}[b]
H_s = (g_x\sigma_x + g_y\sigma_y + g_z\sigma_z)\cos(\omega_s t + \phi_s)
\end{gathered}
\end{equation}

where $g_x, g_y$ and $g_z$ are the signal amplitudes along each axis, $\sigma_i$ are the Pauli operators, $\omega_s$ is the signal frequency and $\phi_s$ is signal phase. We move through two rotating reference frames, first with respect to $\frac{1}{2} \omega_0 \sigma_z$, and then $\frac{1}{2} \omega_m \sigma_x^\prime$ (primes denote the reference frame), and select a single sensor resonance for each axis of propagation;

\begin{equation}
\begin{gathered}[b]
H_{s,x}^{\prime \prime} = \frac{1}{2}g_x\sigma_x^{\prime \prime} \cos((\omega_s - \omega_0)t + \phi_s) 
\label{eq:Hsx}
\end{gathered}
\end{equation}
\begin{equation}
\begin{gathered}[b]
H_{s,y}^{\prime \prime} = - \frac{1}{2}g_y\sigma_y^{\prime \prime} \sin((\omega_s - \omega_0)t + \phi_s)
\label{eq:Hsy}
\end{gathered}
\end{equation}
\begin{equation}
\begin{gathered}[b]
H_{s,z}^{\prime \prime} = \frac{1}{2}g_z\sigma_z^{\prime \prime} (\cos((\omega_s - \omega_m)t + \phi_s) - \sin((\omega_s - \omega_m)t + \phi_s)) 
\label{eq:Hsz}
\end{gathered}
\end{equation}

where $\omega_0$ describes the electron spin resonance and we choose $\omega_m = \Omega$ to meet the conditions of the CCDD scheme. Choosing $\epsilon_m = \omega_s - \omega_0$ for Eqs. \ref{eq:Hsx} and \ref{eq:Hsy}, and $\epsilon_m = \omega_s - \omega_m$ for Eq. \ref{eq:Hsz} satisfies the resonance conditions of the sensor. Our aim is to demonstrate that for each signal in Eqs. \ref{eq:Hsx}, \ref{eq:Hsy} and \ref{eq:Hsz}, the sensor response will depend on the signal phase $\phi_s$. To illustrate this we model the spin evolution for each signal when driven by the CCDD field. Damping effects are not considered. As the sensor readout projects the spin onto the z-axis, in Supplementary Fig. \ref{SIfig4} we plot the z-component of the spin vector as a function of time, for signal phases of $\phi_s = 0, \frac{\pi}{2}$ and signals along the x, y and z axes in (a), (b) and (c) respectively. For each signal, the spin has a phase-dependent response, illustrating the protocols ability to resolve signal phase for any signal vector.

\begin{figure*} 
\centering
\includegraphics[width=0.5\columnwidth]{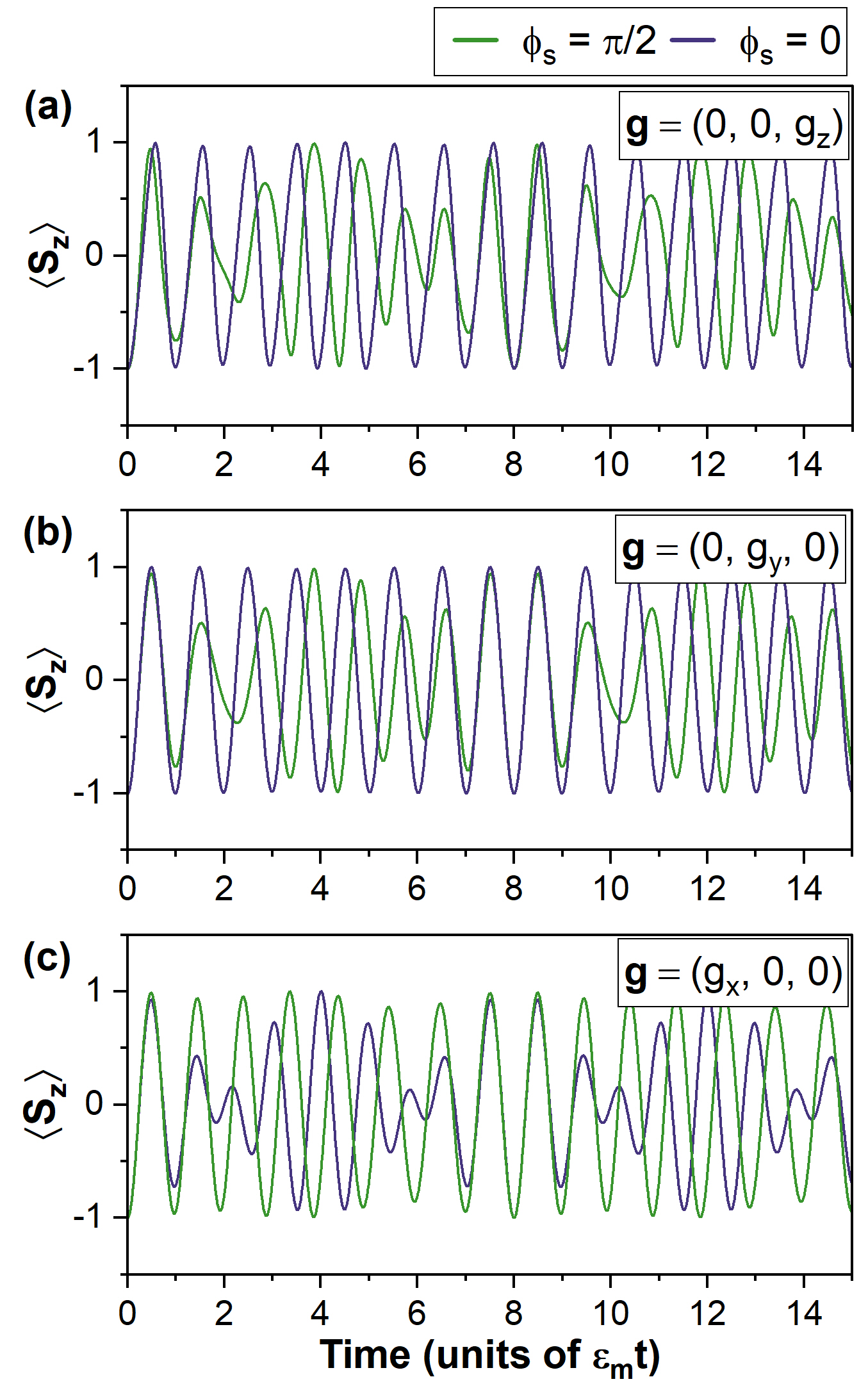}
\caption{\textbf{Rotating frame spin dynamics modelled under different signal vectors.} Z projection of the rotating frame spin vector as a function of time, modelled for a signal vector of \textbf{g} = (0, 0, $g_z$) and signal frequency of $\omega_s = \Omega + \epsilon_m$ in \textbf{(a)}, \textbf{g} = (0,  $g_y$, 0) and signal frequency of $\omega_s = \omega_0 + \epsilon_m$ in \textbf{(b)} and \textbf{g} = ($g_x$,  0, 0) and signal frequency of $\omega_s = \omega_0 + \epsilon_m$ in \textbf{(c)}. Modelled for signal amplitudes of $g_{x, y, z} = \frac{1}{4} \epsilon_m$ and signal phases of $\phi_s = 0$ (blue) and $\phi_s = \frac{\pi}{2}$ (green).}
\label{SIfig4}
\end{figure*}

\subsection{\label{sec:4}{Device Response to Signal Amplitude}}

\begin{figure*}[h!] 
\centering
\includegraphics[width=0.5\columnwidth]{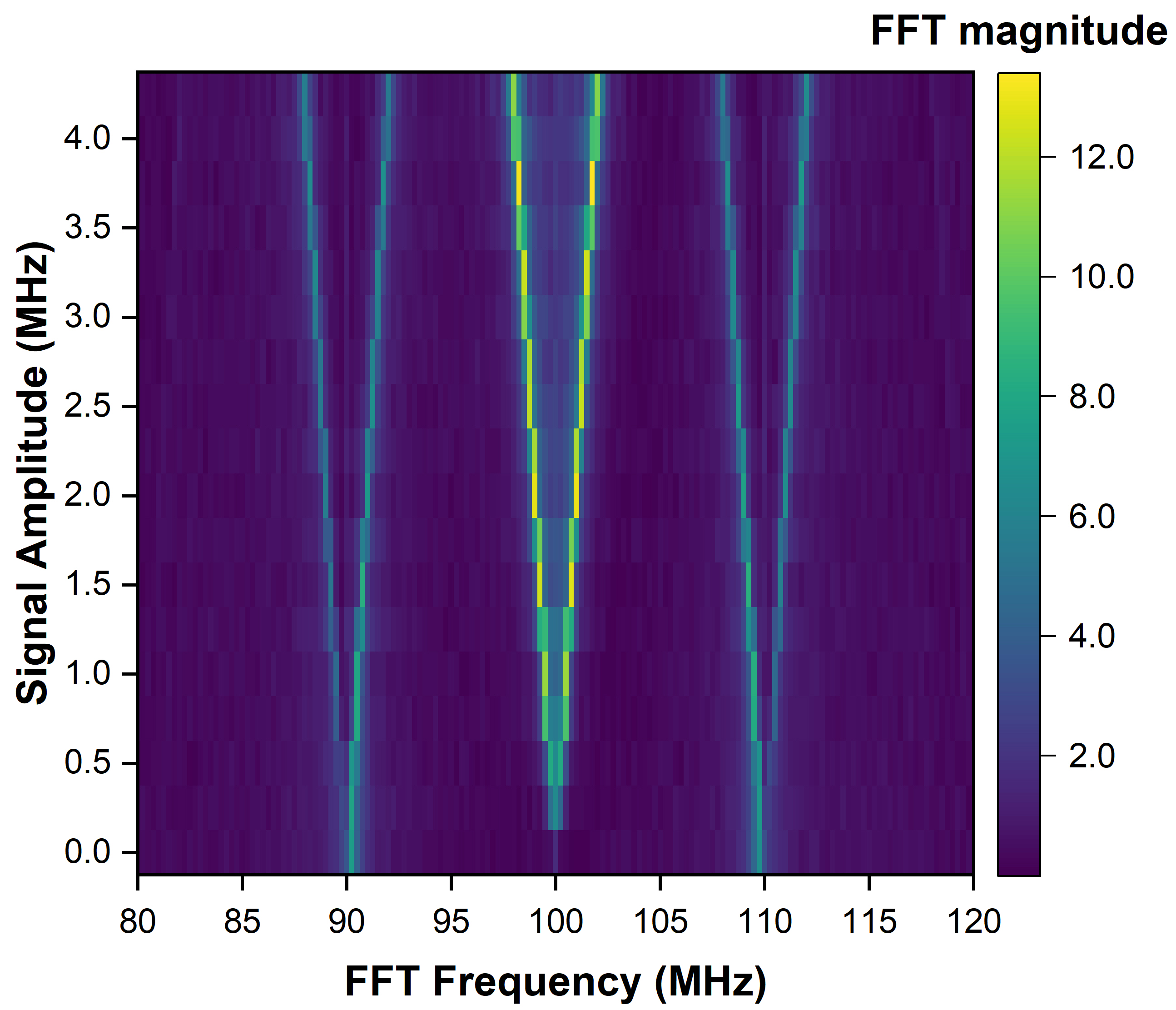}
\caption{\textbf{Fourier response to signal amplitude} Mollow triplet structure as a function of signal amplitude. Each Fourier transform was produced from a Rabi oscillation recorded over 4000 ns, for a fixed signal amplitude. The signal phase was $\phi_s = \frac{\pi}{2}$ at a frequency $\omega_s = 2.31 \: \mathrm{GHz}$. The divergent Fourier components illustrate the devices sensitivity to signal amplitude.}
\label{SIfig5}
\end{figure*}

In the Fourier domain, the frequencies of the nested Mollow triplet structure are determined by the field amplitudes of the CCDD microwave drives, $\Omega$ and $\epsilon_m$, and the signal, $g_x$. This means that, for a fixed set of CCDD amplitudes, small changes in the signal amplitude $g_x$ will change the Rabi frequencies of the sensor. Monitoring the contrast at a fixed point in the Rabi oscillation provides a way of detecting any changes to the Rabi frequency - and thereby the signal amplitude. If the fixed point is selected appropriately, changes in signal phase and amplitude produce opposite responses, such that changes in the two can be distinguished (see Fig. 3(a) of the main text). Supplementary Fig. \ref{SIfig5} plots the Mollow triplet structure as a function of signal amplitude, for a signal phase of $\phi_s = \frac{\pi}{2}$. The Fourier components centre on the CCDD drive fields $\Omega = 100\: \mathrm{MHz}$ and $\Omega \pm \epsilon_m = 100 \pm 10 \: \mathrm{MHz}$. The significant Fourier components are offset from these central values by the signal amplitude, to $\Omega \pm g_x = 100 \pm g_x \: \mathrm{MHz}$ and $\Omega \pm \epsilon_m \pm g_x= 100 \pm 10 \pm g_x \: \mathrm{MHz}$. This manifests as diverging Fourier components in Supplementary Fig. \ref{SIfig5}, illustrating the devices sensitivity to signal amplitude.

\subsection{\label{sec:5}{Optimised Heterodyne Measurement Time}}

The quantum heterodyne detection protocol presented in the main text tracked the evolution of signal phase across successive readouts, with each lasting 5 $\mathrm{\mu s}$. Note that its beneficial to minimise the length of this readout sequence. Firstly, it enables a higher sampling resolution, which improves the SNR and increases the Fourier resolution. It also increases the Nyquist frequency of the heterodyne protocol, increasing the frequency range that can be detected for a fixed set of CCDD parameters. In our experiment this was constrained by the hardware, as the timing was maintained using an arbitrary waveform generator (AWG) with a memory of 1 ms. To use the CCDD as a coherent phase reference over longer acquisition times, an integer multiple of readouts therefore needed to be applied within each 1 ms window. 

The 5 $\mathrm{\mu s}$ sequences used in the main text consisted of 2 $\mathrm{\mu s}$ of optical initialisation/ readout, 950 ns of CCDD drive and 2050 ns of idle time. This relatively long CCDD drive time maximised the amplitude sensitivity, as it provided more time for weak signals to drive a detectable change in spin state. This can be dramatically shortened if amplitude detection is not required, however. To demonstrate phase sensitivity its sufficient to differentiate between two Fourier responses which are largely defined by the drive amplitude $\epsilon_m$ for $\epsilon_m >> g_x$. This is illustrated by the Rabi sequence and associated Fourier transform in Supplementary Figs. \ref{SIfig6}(a) and (b), where the inset shows a departure in the sensors phase response on timescales of $\approx 1/ (\epsilon_m - g_x) \approx 125 ns$. Results presented here used $\epsilon_m = 10 \mathrm{MHz}$, however the CCDD drive functions with values of $\epsilon_m$ up to 70 MHz \cite{Ramsay2023}, which could further reduce this time to $\approx 1/ (\epsilon_m - g_x) \approx 15$ ns. A single readout sequence could therefore be reduced to 2.5 $\mathrm{\mu s}$ using the same 2 $\mathrm{\mu s}$ optical initialisation/ readout time used previously. There is also scope to reduce this however, by increasing the laser power or using a higher power microscope objective, potentially reducing the readout time to 1.25 $\mathrm{\mu s}$.

\begin{figure*}[h!] 
\centering
\includegraphics[width=1\columnwidth]{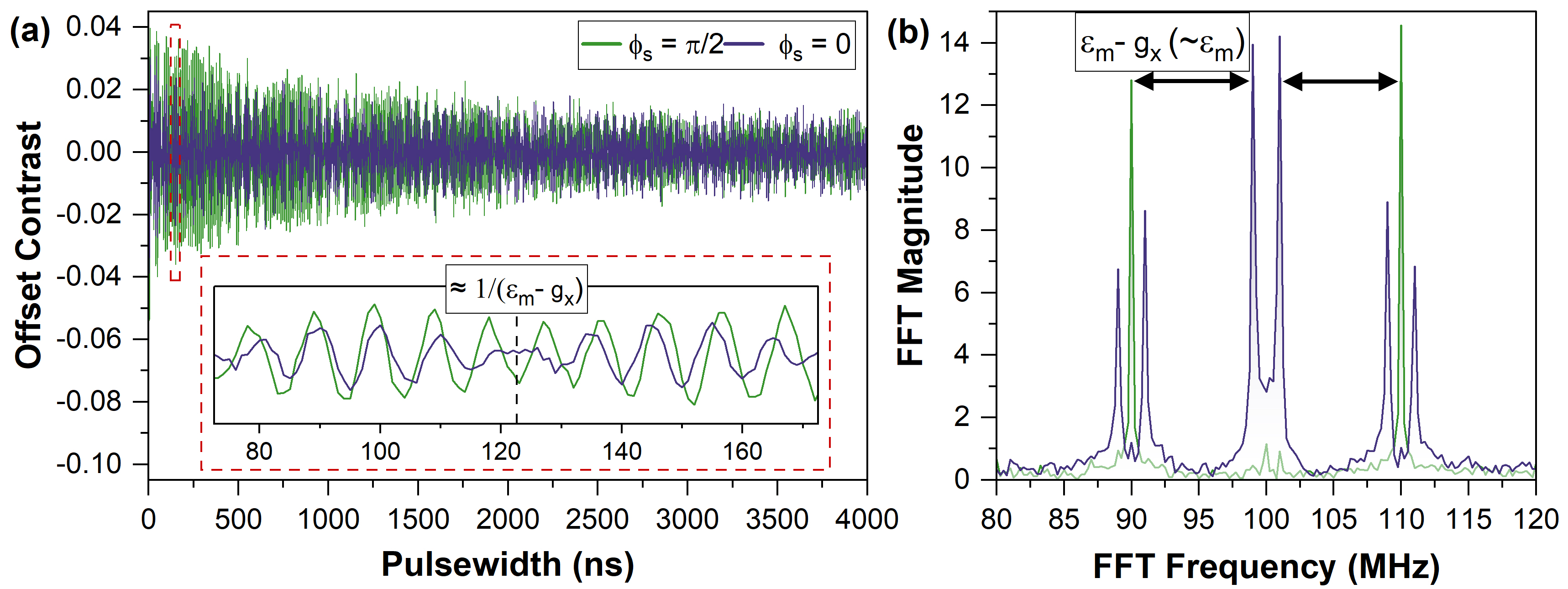}
\caption{\textbf{Optimisation of heterodyne measurement time} \textbf{(a)} CCDD Rabi oscillation exposed to a resonant signal with phases of $\phi_s = 0$ (blue) and $\phi_s = \frac{\pi}{2}$ (green). The sensor can differentiate between signal phases on timescales of $\approx 1/(\epsilon_m - g_x)$, shown in the inset. \textbf{(b)} Fourier transform of (a). Each signal phase produces a different Fourier response. For small signal amplitudes where $g_x << \epsilon_m$, the major Fourier components contributing to each response are largely determined by the drive parameter $\epsilon_m$.}
\label{SIfig6}
\end{figure*}

\providecommand{\noopsort}[1]{}\providecommand{\singleletter}[1]{#1}%